\documentclass{aa}
\usepackage{graphicx}
\newcommand{\on}{\textsc{on}}
\newcommand{\off}{\textsc{off}}

\newcommand{\cerenkov}{\v{C}erenkov}

\newcommand{\gmin}{$\gamma\ \textrm{min}^{-1}$}
\newcommand{\xr}{X-ray}

\newcommand{\gr}{$\gamma$-ray}
\newcommand{\grs}{$\gamma$-rays}
\newcommand{\vhe}{V\textsc{HE}}

\newcommand{\dg}{\ensuremath{^\circ}}

\def\ltsim{\ensuremath{\hbox{\rlap{\hbox{\lower4pt\hbox{$\sim$}}}\hbox{$<$}}}}
\def\gtsim{\ensuremath{\hbox{\rlap{\hbox{\lower4pt\hbox{$\sim$}}}\hbox{$>$}}}}

\newcommand{\mfv}{Mkn~501}
\newcommand{\mfr}{Mkn~421}
\newcommand{\bllac}{BL~Lac}
\newcommand{\bllacs}{BL~Lacs}
\newcommand{\source}{1ES~1426+428}
\newcommand{\cat}{C\textsc{at}}

\newcommand{\hegra}{H\textsc{egra}}
\newcommand{\beppo}{BeppoS{\textsc{ax}}}

\newcommand{\egret}{E\textsc{gret}}

\newcommand{\mjd}{MJD}

\begin{document}
   \title{Detection of the {\bllac} object {\source} in the Very High
   Energy gamma-ray band by the CAT Telescope from 1998-2000}


   \author{Arache~Djannati-Ata\"\i \inst{1}\and
B.~Khelifi \inst{1} \and
S.~Vorobiov \inst{2}\and
R.~Bazer-Bachi \inst{3} \and
G.~Debiais \inst{4} \and
B.~Degrange \inst{2} \and\\
P.~Espigat \inst{1} \and
B.~Fabre \inst{4} \and
G.~Fontaine \inst{2} \and
P.~Goret \inst{5} \and
C.~Gouiffes \inst{5} \and
C.~Masterson \inst{1} \fnmsep\thanks{\emph{Present address:} MPI
Kernphysik, Heidelberg, Germany}\and
F.~Piron \inst{2}  \fnmsep\thanks{\emph{Present address:} GAM,
Univ. de Montpellier~II, France} \and \\
M.~Punch \inst{1} \and
M.~Rivoal \inst{6} \and
L.~Rob \inst{7} \and
J.-P.~Tavernet \inst{6}
}

\institute{
Physique Corpuscul\-aire et Cosmologie, Coll{\`e}ge de France et
Universit{\'e} Paris VII, France (IN2P3/CNRS) 
\and
Laboratoire Leprince-Ringuet,
Ecole Polytechnique, Palaiseau, France (IN2P3/CNRS)
\and
Centre d'Etudes Spatiales des Rayonnements, Universit{\'e} Paul
Sabatier, Toulouse, France (INSU/CNRS) 
\and
Groupe de Physique Fondamentale, Universit{\'e} de Perpignan, France
\and
Service d'Astrophysique, Centre d'Etudes de Saclay, France (CEA/DSM/DAPNIA)
\and
Laboratoire de Physique Nucl{\'e}aire et de Hautes Energies,
Universit{\'e}s Paris VI/VII, France (IN2P3/CNRS)
\and
Institute of Particle and Nuclear Physics, Charles University, Prague, Czech Republic
}

   \mail{djannati@in2p3.fr}
   \offprints{A.~Djannati-Ata\"\i} 

   \date{Received xxxxxx nn, 2002; accepted xxxxxx nn, 2002}

   \abstract{The {\bllac} Object {\source}, at a red-shift of
$z$$=$$0.129$, has been monitored by the {\cat}
telescope from February 1998 to June 2000. The accumulation of 26 
hours of observations shows a {\gr} signal of 321 events above
$250\:\mathrm{GeV}$ at 5.2 standard deviations,
determined using data analysis cuts adapted to a weak, steep-spectrum
source. 
The source emission has an average flux of  
${\Phi}_\mathrm{diff}({400\:\mathrm{GeV}})=6.73\pm1.27^\mathrm{stat}\pm1.45^\mathrm{syst}\times10^{-11}\:\mathrm{cm^{-2}\:s^{-1}TeV^{-1}}$, 
and a very steep spectrum, with a differential spectral index of
$\gamma$$=$$-3.60$$ \pm $$0.57$ which can be refined to
$\gamma$$=$$-3.66$$ \pm $$0.41$ using a higher flux data subset.   
If, as expected from its broad-band properties, the Very High Energy
emission is hard at the source, these observations support a strong
absorption effect of {\grs} by the Intergalactic Infrared field.

   }
\titlerunning{Detection of 1ES~1426+428 by \cat\ from 1998-2000 observations}

   \maketitle
%
\section{Introduction}
{\bllac} objects, together with Flat Spectrum Radio Quasars (FSRQs),
constitute the extreme class of Active Galactic
Nuclei known as blazars. They exhibit a highly variable non-thermal
spectral energy distribution (SED) which is interpreted as the Doppler
boosted emission of a jet pointing towards the Earth.

While at least 66 blazars are reported in the third {\egret} catalogue
(\cite{Hartman99}), in the TeV range only two
nearby {\bllac} objects, {\mfr} ($z$$=$$0.031$) and {\mfv} ($z$$=$$0.034$),
have been firmly established so far. This can be explained partially by
the small field of view of ground-based telescopes (a few degrees),
although, as compared to satellite-borne instruments, they benefit from a
much larger sensitivity.
The attenuation of \grs\ through pair-production with Intergalactic Infrared
field (IIR) is another limiting factor for viewing distant sources in
the Very High Energy ({\vhe}) domain. However, the detection of
absorption features in blazar spectra can provide interesting
constraints on the poorly-measured $0.5-20\:\mu\mathrm{m}$ IR band.

In this context, a survey of nearby {\bllacs} has been carried out with
the {\cat} telescope since 1997. Among them, {\source}
($\alpha_{J2000}$$=$$\mathrm{14^h28^m32.7^s}$, 
$\delta_{J2000}$$=$$+42\dg40'21''$) at a red-shift of
$z=0.129$,  
occupies a peculiar position: the characteristics of its radio-to-\xr\
SED are very
similar to those of the two TeV 
{\bllacs}, especially its hard \xr\ emission which flags the presence
of ultra-relativistic electrons. 
On the other hand, the relatively high red-shift of \source\ 
can imply significant absorption by the IIR photons.

VHE \gr\ detections of  {\source} have recently been reported by
Whipple and {\hegra} during 1999-2001 (\cite{Horan02}, \cite{Aharonian02}).
In this letter, we report on its detection based on
observations made by the \cat\ telescope from February 1998 to June
2000. In Sect. 5 we will also give a first estimation of its
spectrum.

\section{The CAT telescope and its standard data analysis}
The \cat\ 
imaging \cerenkov\ telescope, equipped with a $3.1\dg$ high
resolution camera ($4.8\dg$ full field of view) and a $17.8\:\mathrm{m}^2$  
mirror, is located in the French Pyr{\'e}n{\'e}es at an altitude of 1650 m 
a.s.l. (\cite{Catdet98}). The instrument records \cerenkov\ light from
the particles in  
cosmic-ray showers. The analysis is based on the comparison
of individual events and theoretical average images of \grs\ as a
function of impact parameter and energy (\cite{LeBohec98}).
To separate \gr\ initiated shower images from those due to
hadronic showers the $\chi^2$-like goodness-of-fit parameter,
$P_{\chi^2}$, given by the above procedure is used, together with a
cut on the pointing angle, $\alpha$ --- which is the angle at 
the image barycentre between the actual source position and the
reconstructed image axis of the \gr\ candidate --- while requiring a
minimum total charge, $Q_{\rm p.e}$ in photo-electrons (p.e.). The
standard cuts are dedicated to
strong and fairly hard sources (``SH'' hereafter). They are 
$P_{\chi^2}\:$$>\:$$0.35$, $\alpha\:$$<\:$$6\dg$ and $Q_{\rm
p.e}\:$$>\:$$30\:\mathrm{p.e.}$ 
(\cite{Piron01}), yielding a sensitivity of 4.5 standard deviations
($\sigma$) per $\sqrt {T_{\rm ON}/1{\rm h}}$ for a Crab-like source at
transit ($\sim$$21$$\dg$ from Zenith) and a rate of 1.8 {\gmin} after cuts.

\section{Weak point source data analysis}
Given the relatively high red-shift ($z$$=$$0.129$) of {\source} as
compared to the two confirmed \bllacs, {\mfr} and {\mfv}, one
expects a rather weak flux and a soft
spectrum due mainly to source distance and the IIR attenuation effect.   
The standard cuts have therefore been modified, trading-off \gr\
efficiency against background rejection.  
A study was made to get an indication of the
IIR softening effect. Many estimates and models have been proposed for
the IIR density (see \cite{Hauser01} for a review).
Conservative assumptions yield roughly a change of $\sim 1.0-1.5 $ of the
spectral index for the red-shift considered here. On the other hand, given its 
hard \xr\ emission, a rather hard \vhe\ spectrum is expected at the
source, i.e. $\sim -2$. 
Consequently a weak simulated signal 
(1~\gmin\ before cuts) with a spectral index $\sim -3$ was tested against
real background data taken 
from \off\ observations of other sources (with roughly the same elevation as
{\source}).
For such a steep spectrum, a tighter cut on $P_{\chi^2}$ and  a
higher value of $Q_{\rm p.e}$ were found to improve the background
rejection, while a looser cut on $\alpha$ improves the efficiency because
of the low-energy nature of the signal (for low energy \grs\ the
estimation of the direction is less precise).

\begin{table}[h!]
  \begin{center}
    \caption{ Cut values and results for MC generated signals for a
    source with a hard spectrum, after simulated IIR absorption.  
    The first and second lines show results for the SH and WS cuts,
    respectively. $\epsilon_{\gamma}$ is the \gr\ efficiency,
    $R_\mathrm{h}$ is the rejection factor for background cosmic rays,
    $QF$ is the quality factor defined as $QF=\epsilon_{\gamma} \times
    \sqrt R_{h}$. The final column gives the rate obtained on  
      real Crab data in {\gmin}. Its variation from SH to WS is not
    the same as that of $\epsilon_{\gamma}$, due to the different spectral
    shape of the MC signal as compared to the Crab.
}
    \begin{tabular}{c | c c c | c c c | c}
      \hline
Cuts  &$P_{\chi^2}$ &$\alpha$ &$Q_{\rm p.e}$ &$\epsilon_{\gamma}$  &$R_{h}$
  &$QF$  &{\gmin} \\ 
  &min &max &min & & & &Crab\\
      \hline 
SH   &0.35 & 6\dg\ &30  &32\%           &180      &$4.3$    &1.8\\ 
WS   &0.50 & 8\dg\ &45  &25\%           &480      &$5.5$    &1.15\\ 
      \hline
      \end{tabular}
    \label{TabCuts}
  \end{center}
\end{table}

The final values of the cuts (``WS'' for Weak and
Soft) were chosen by requiring a minimum efficiency 
 of 25 \% such as to keep a sufficient number of \gr\ events for
 spectral measurements and to avoid systematic effects due
 to a too small efficiency : $P_{\chi^2}$$>$$0.5$,  
 $\alpha$$<$$8$$\dg$ and $Q_{\rm p.e}$$>$$45$ p.e. The resulting quality 
 factor, $QF$, and \gr\ rate on the Crab nebula, as
 compared to SH cuts, are given in Table~\ref{TabCuts}. It has been 
verified that using the WS cuts
for spectral analysis of Crab data yield, as
expected, the same results as obtained by the standard cuts. 

\section{Data sample and detected signal}

\begin{table}[b!]
  \begin{center}
    \caption{Selected data and results for the three observing
    seasons from 1998 to 2000. The final column gives the significance
    of each detection according to the maximum likelihood definition
    by \cite{LietMa83}.
    }
    \begin{tabular}{l r l r c c }
      \hline
	Year  	&\on\ (h) 	&\off\ (h)  	& Excess & \gmin\ & $\sigma$\\
      \hline
	1998	&8.9		&24.7 		&  91 	&0.17	&3.1 \\
	1999	&13.4  		&30.5 		& 186 &0.25	&3.9 \\
	2000	&3.7   		&17.2   	&  44 &0.20	&1.8 \\
        \hline
	Total	&26.0   	&72.4   	& 321  &0.21	&5.2 \\
        \hline
      \end{tabular}
    \label{TabSignal}
  \end{center}
\end{table}

The observations reported in this paper were made with the \cat\ telescope
from February 1998 to June 2000, in the \on-\off\ mode, where
control regions with the same trajectory on the sky are used as 
off-source data (i.e., with the same declination and shifted 
right ascension).

\begin{figure}[tb!]
\begin{center}
\hbox{
 \includegraphics[width=0.99\linewidth,height=7.5cm]{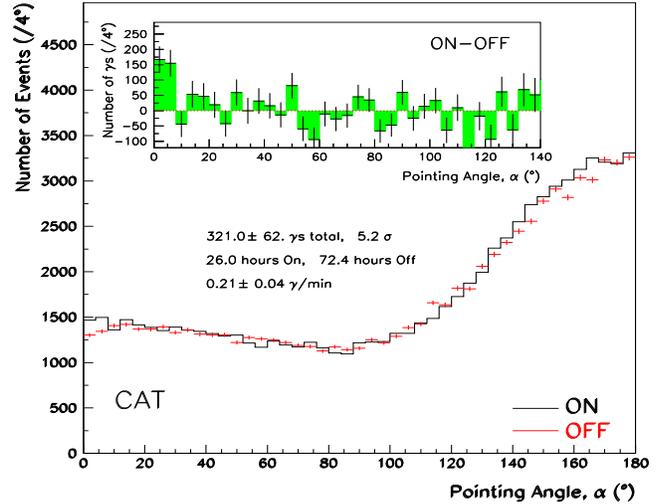}     
}
\caption[]{ The accumulated \on -\off\ pointing angle $\alpha$ plot
for the three observing periods from 1998-2000. The \off\
  normalization has been done using the 20--100\dg\ control region.
The increasing number of events at large $\alpha$ ($>$$100$$\dg$)
corresponds to images cut by the camera border: the slight  
difference between \on\ and \off\ samples in this region is a
systematic edge effect which is not relevant to the signal or spectrum
evaluations.
} 
\label{AlfaPlots}
\end{center}
\end{figure}


A set of 26 hours \on\ and  12.5 hours \off\, was selected according
to criteria for 
good weather and smooth telescope operation (\cite{PironThese}). Using
the  \cat\ data-base, the \off\ sample was expanded by adding 59.9 hours
of observations of control regions (\off) made during the same 
observing periods 
and with the same zenith angle range (i.e. 0--32{\dg}). The
compatibility of the two \off\ 
sample sets was checked over the tracking ratio distributions
(i.e., the ratio of events within the $\alpha$ cut to the 20--100\dg\
control region) for all energy and zenith angle bins. This is the only
relevant distribution to compare for signal and spectrum calculations. 
Figure~\ref{AlfaPlots} shows the accumulated \on-\off\ pointing angle
$\alpha$ plot. An excess of 321 \gr\ events,
corresponding to a mean rate of 0.21 {\gmin}, is observed
with a significance level of 5.2 $\sigma$.
The measured \gr\ rate is 18\% of
that of the Crab nebula within the same cuts (see Table~\ref{TabCuts}).
There is a significant sensitivity improvement with the WS set as
compared to SH cuts: the latter
yield an excess of 367 events at 3.8 $\sigma$.

The detected signal in each observing season is summarized in
Table~\ref{TabSignal}. In 1998, an excess of $3.1\sigma$  was seen in
8.9 hours (February--May) with no flaring period evident.

\begin{figure}[tbh]
\begin{center}
\hbox{
\includegraphics[width=0.99\linewidth,height=7cm]{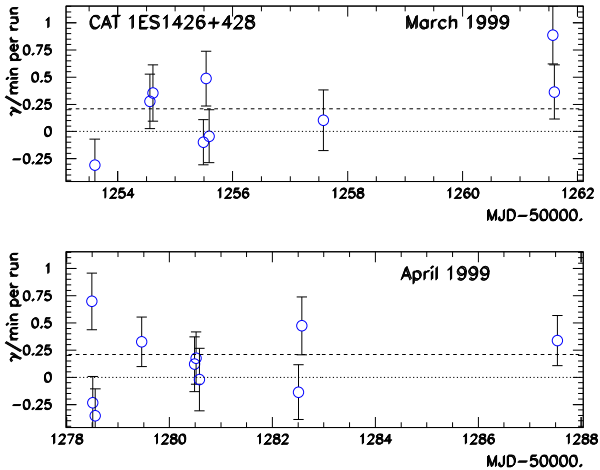}
}
\caption{Event rate of {\source} during March and April
1999 in {\gmin}.The source flux was the highest during \mjd\ 51261.5
(0.8 Crab, $3.4\sigma$) and \mjd\ 51278.5 (0.6 Crab, $2.7\sigma$). The
dashed line shows the average emission level.
}
\label{PlotLC99} 
\end{center}
\end{figure}

From February to June 1999, a $3.9\sigma$ signal was
detected in 13.4 hours. This period was consecutive to the campaign by
\beppo\ in February 1999 
where the \xr\ flux was very low as compared to archive data, but
showed an extremely hard \xr\ spectrum with a differential index of $-0.92$
(\cite{Costamante01}). The source flux was the highest during two
nights in  March and April with some indication of rapid variability
(see Fig.~\ref{PlotLC99}): 
 a rate of 0.8 Crab ($3.4\sigma$) followed by 0.3 Crab
($1.5\sigma$) was detected in two runs (each of 30 min) at \mjd\ 51261.5
(23-24/03/99). During \mjd\ 51278.5 (9-10/04/99) a rate of 0.6 Crab
($2.7\sigma$) was followed immediately by two runs showing no signal.


Weather conditions in the Pyr{\'e}n{\'e}es limited the observations in the
\on\ - \off\ mode to only 3.7 hours during spring 2000 (some other
data are available in 2000, but with a different observation mode
which is not analysed here). The measured \gr\ rate is nevertheless
compatible with that of the preceding periods.

\section{Spectral measurements}

{\cat}'s spectrum analysis method has been applied to the data.
The method (\cite{Djannati99,Piron01}) relies on a maximum-likelihood
estimation of the parameters taking into account 
the Poissonian probability distributions of the \gr\ and background
events, as well as the instrument response functions (effective area and
resolution) derived by Monte-Carlo simulations and validated by
comparison against nearly pure \gr\ samples (\cite{Piron99}). 
Data are classified within zenith angle and energy bins. 
The high energy bins show no excess:  when requiring the energy per
event to be less than $1.3\:\mathrm{TeV}$ the significance level rises to 
6.1$\sigma$ while almost no variation in the total number of 
excess events is seen.    
Consequently, a power-law is adjusted in the limited range of
$250\:\mathrm{GeV}$ to $1\:\mathrm{TeV}$, 
yielding a differential flux 
$\phi_0({400\:\mathrm{GeV}})=6.73\pm1.27^\mathrm{stat}$
in $10^{-11}\:\mathrm{cm^{-2}s^{-1}TeV^{-1}}$
units, and a very steep differential index $\gamma=-3.60$$\pm$$0.57$.
With no cut on the maximum energy (i.e., from $250\:\mathrm{GeV}$ to
$10\:\mathrm{TeV}$) the fitted value of the index increases:
$\gamma=-4.04\pm0.52$. The corresponding integral flux has been
computed above the energy for which the 
errors on the two parameters $\phi_0$ and $\gamma$ de-correlate:
$\phi_\mathrm{int(>300\:\mathrm{GeV})}=2.17\pm0.39 \times 
10^{-11}\:\mathrm{cm^{-2}s^{-1}}$.

\begin{figure}[t!]
\begin{center}
\hbox{
 \includegraphics[width=0.99\linewidth,height=7.5cm]{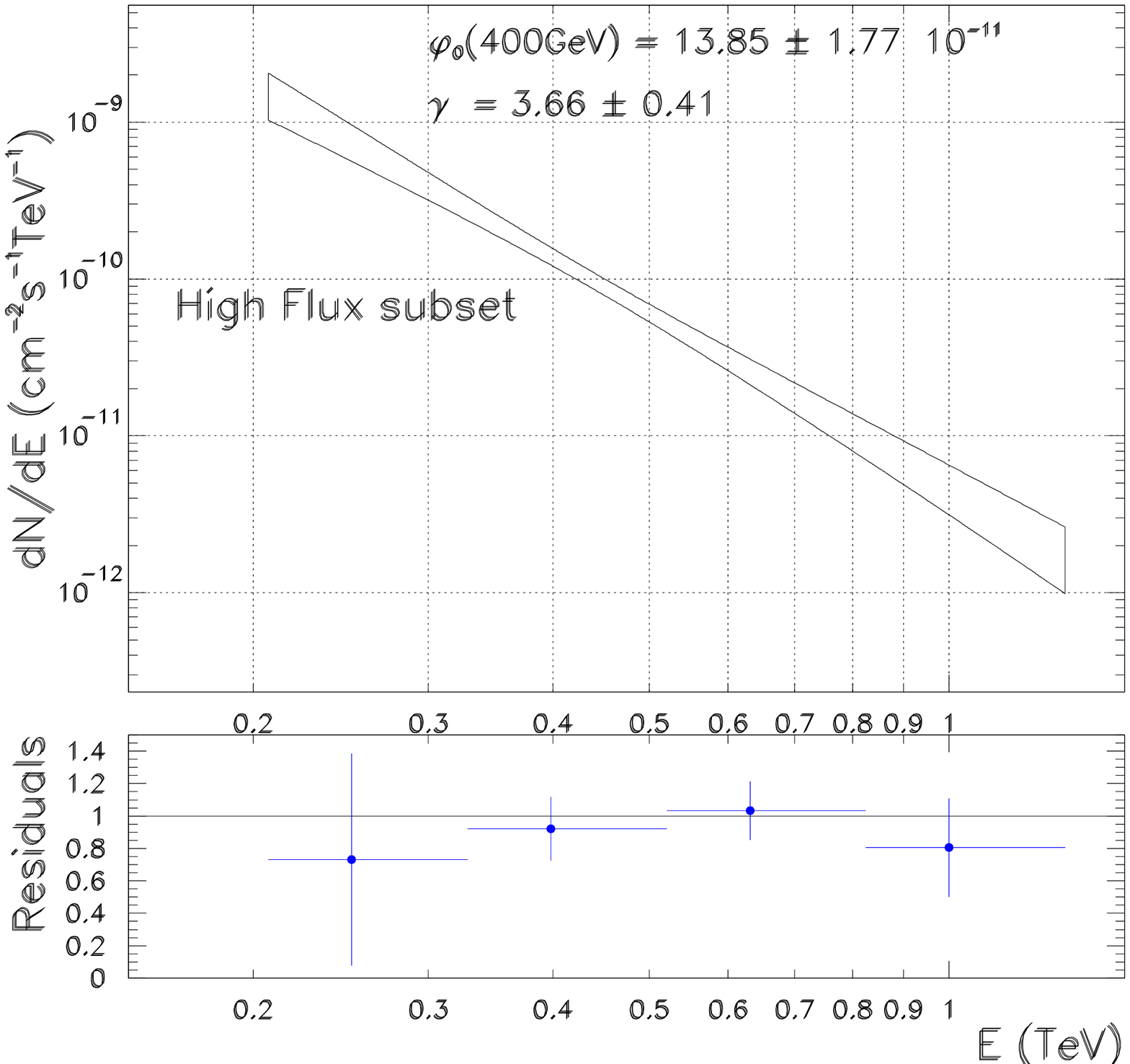}     
}
\caption[]{
{\source} time-averaged spectra between $0.25$ and $1\:\mathrm{TeV}$
from 1998 to 2000, for a power-law hypothesis using a High Flux subset with
a  mean \gr\ rate of 0.5 {\gmin}.  
The areas show the $68$\% confidence-level contour given by the
likelihood fit:  
$\phi_0=13.85\pm1.77^\mathrm{stat}\pm2.77^\mathrm{syst}
\times (E/{\mathrm{400 GeV}})^{-3.66\pm0.41^\mathrm{stat}}$
in $10^{-11}\:\mathrm{cm^{-2}s^{-1}TeV^{-1}}$ units.

The lower plot gives the ratio, in each bin of {\it estimated} energy
(as opposed to true energy bins in the upper plot),
of the predicted number of \gr\ events 
to the observed signal. The energy at which the errors on flux and
index de-correlate is $430\:\mathrm{GeV}$.
} 
\label{SpecPlot}
\end{center}
\vspace{-0.6cm}
\end{figure}

To get a better determination of the spectral index, one can
improve the  signal-to-background ratio by selecting the data,
requiring a minimum \gr\ rate (High Flux subset). We make here the
reasonable assumption 
that the background spectrum variations as a function of the
statistical fluctuations of the tracking ratio are negligible. On the other
hand, if the intensity-hardness correlations 
seen on other {\bllac} sources (\cite{Djannati99,Aharonian01}) hold for
{\source}, the selection could in principle bias the time-averaged
spectrum.  
However, given the weakness of the signal and the magnitude of a
possible spectral variability ($<0.5$) as compared to that of the
statistical error, the bias should be negligible here. 
In consequence, a subset of 10.7 hours of {\on} data with a minimal \gr\ rate
of 0.2 {\gmin} was selected (corresponding to an average rate of  0.5
{\gmin}), and a differential spectral 
index $\gamma=3.66\pm0.41$ was derived.
Figure~\ref{SpecPlot} shows the $68$\% confidence-level contour given by the
likelihood estimator as well as the residuals to the fit for this subset.

Systematic errors could arise from the uncertainty on the
absolute energy scale (variations of the atmosphere transparency and
light-collection efficiencies during the observation periods) and, to
a lesser extent, from limited Monte-Carlo statistics in the
determination of the response functions of the instrument. They have been
estimated from detailed simulations (\cite{PironThese}):  
$(\Delta\phi_0/\phi_0)^\mathrm{syst}=\pm20$\% and
$(\Delta\gamma)^\mathrm{syst}=\pm 0.06$.
Here, the systematic error in the index remains 
negligible compared to the statistical error, while for the flux
these errors are comparable in magnitude.

\section{Discussion}

Observations of \source\ from 1998 to 2000 by \cat\ show a signal at
$5.2\sigma$ with an average event rate of order of $\sim 0.2$
Crab. This is the most distant detected \bllac\ in the \vhe\ range.

For the three observing seasons studied here the source average
emission seems stable, within the measurements'
accuracy, while there is some evidence for time variability with transient
emission rates $> 0.5$ Crab. 

Given its broad-band properties, \source\ was one of the most promising \vhe\
candidates, despite its red-shift of
$>\:0.1$  (\cite{Costamante99}). Its detection supports    
the unifying scheme by \cite{Ghisellini98}, where the lower
luminosity blazars with high frequency peaked synchrotron emission
(HBLs) are efficient accelerators to very high energies. 
It is remarkable that \xr\ measurements
(\cite{Costamante01}), contemporaneous to \cat\ observations in
February 1999 reported a peak emission energy of $\sim 100\:\mathrm{keV}$,
comparable to that of {\mfv} during its strong activity in 1997. 

If, as suggested by these data, the same acceleration/cooling mechanisms
are at play within the 
two sources under similar conditions, one would expect a hard \vhe\
differential spectrum with 
$\gamma \sim 2.0$ at the source (see \cite{Djannati99}). The very
steep spectrum observed here, $\gamma=3.66\pm0.41$,
supports a strong absorption of {\grs} by the diffuse
Intergalactic Infrared field, even though the effect is compatible
with several --- among many --- estimates and models of its density in the 
$0.5-20\:\mu\mathrm{m}$ band.  Any
definite conclusion on this subject and any measurement of the IIR
density will require the detection of more sources at different
red-shifts. 

\begin{acknowledgements}
The authors wish to thank Electricit{\'e} de France for making
available to them equipment at the former solar plant ``Th{\'e}mis''.
L.R. thanks for the financial support granted by IPNP Project MSMT
(LN00A006).  
\end{acknowledgements}

\end{document}